\documentclass[10pt,a4paper]{article}

\usepackage{epsfig,amsfonts,amssymb,amsmath,graphicx}

\begin{document}

\begin{flushright}
{\small LPTh-Ji 09/002}
\end{flushright}

\begin{center}
\bigskip\vspace{1.8cm}

{\Large \textbf{Sphalerons on Orbifolds}}\vspace{1.3cm}

\textbf{Amine Ahriche}\\[0pt]

\bigskip

\textit{LPTh, University of Jijel, PB 98, Ouled Aissa, DZ-18000 Jijel,
Algeria.\\[0pt]LPMPS, University of Constantine, Ain El-Bey, DZ-25000
Constantine, Algeria. \\[0pt]Faculty of Physics, University of Bielefeld,
Postfach 100131, D-33501 Bielefeld, Germany.}
\end{center}

\vspace{1.4cm}

\hrule \vspace{0.5cm} {\normalsize \textbf{{\large {Abstract}}}}

\vspace{0.4cm}

In this work, we study the electroweak sphalerons in a \textit{5D} background,
where the fifth dimension lies on an interval. We consider two specific cases:
flat space-time and the anti-de Sitter space-time compactified on $S^{1}%
/Z_{2}$. In our work, we take the $SU(2)$ gauge-Higgs model, where the gauge
fields reside in the \textit{5D} bulk; but the Higgs doublet is confined in
one brane. We find that the results in this model are close to those of the
\textit{4D} Standard Model (\textit{SM}). The existence of the warp effect, as
well as the heaviness of the gauge Kaluza-Klein modes make the results
extremely close to the \textit{SM} ones.

\vspace{0.8cm} \textbf{Keywords}: Sphalerons, Kalauza-Klein modes,
Warp Factor. \vspace{0.4cm} \hrule \vspace{1.5cm}

\section{Introduction}

The Standard Model (\textit{SM}) of the electroweak and strong interactions
has been very successful in describing nature at energies around the
electroweak scale ($\sim100~$\textit{GeV}). However, it fails in answering
many fundamental questions in particle physics, like, e.g., the hierarchy
problem and the neutrino mass and its smallness, as well other problems
related to cosmology like the baryon asymmetry in the universe and dark
matter. Therefore a more fundamental theory, which describes nature at higher
scales, needs to become known to explain the problems of particle physics and
related topics.

It has been realized that the hierarchy problem could be a consequence of the
existence of extra dimensions \cite{ADD}. A popular realization of this
concept is the so-called Randall-Sundrum model \cite{RS}. There are several
variants of this scenario, depending on whether the extra dimension is finite
(RS1) or infinite (RS2), and on which of the fields is confined in a brane or
lying on the bulk. In the RS1 models, the space-time has the \textit{5D}
anti-de Sitter ($AdS_{5}$) geometry%
\begin{align}
ds^{2}  &  =g_{MN}dx^{M}dx^{N}=a^{2}\left(  y\right)  \eta_{\mu\nu}dx^{\mu
}dx^{\nu}-dy^{2}\\
&  =e^{-2ky}\eta_{\mu\nu}dx^{\mu}dx^{\nu}-dy^{2}, \label{5D}%
\end{align}
where $y$ is the fifth dimension that has the properties $y\equiv y+2\pi R$;
and $y\equiv-y$; it is compactified on a half-circle $S^{1}/Z_{2}$ with two
\textit{4D} boundaries ($y=0,\pi R$). The metric $\eta_{\mu\nu}%
=diag(-1,1,1,1)$ is the usual 4-dimensional one; and $k$ is the $AdS_{5}$
curvature. In this model, the relation between the \textit{Planck} and the
\textit{TeV} scales seems to be natural, \textit{TeV}$\sim w^{-1}%
M_{Pl}=e^{-\pi kR}M_{Pl}$, where the two fixed points of the fifth dimension
$y=0,\pi R$ represent the \textit{Planck} and the \textit{TeV} branes, respectively.

In the first paper \cite{RS}, only gravity resides in the \textit{5D} bulk,
while the \textit{SM} fields are confined in the \textit{TeV} brane. But
problems with some of the \textit{SM}\ fields are that propagating in the bulk
were also considered, like the case of gauge fields \cite{GB1,GB2}, scalars
\cite{GW}, fermions \cite{BF}, the whole \textit{SM} content \cite{BSM}; and
even supersymmetry \cite{Bsusy}.

As mentioned above, the \textit{SM} fails to explain the origin of matter in
the universe \cite{KRS}, it does not fulfill the second and the third Sakharov
criteria for baryogenesis \cite{Sak}. Although, the first criterion, baryon
number violation, is achieved through the \textit{B+L} anomaly \cite{ABJ},
where both of the baryon and lepton numbers are violated by 3 units due to the
possible transition between two equivalent neighboring vacua of the nontrivial
topology of the \textit{SU(2)} model. It was shown\ \cite{ABJ} that this
transition probability is extremely suppressed, $\sim10^{-162}$, but this is
not the case at higher temperatures. The rate of $B$ violating processes is
proportional to $T^{4}$ at the symmetric phase \cite{Tsps} and suppressed like
$e^{-E_{Sp}/T}$\ in the broken phase \cite{Tspb}, where $E_{Sp}$\ is the
system's static energy within the so-called sphaleron configuration
\cite{Man0,Man}; a field configuration that corresponds to the top of the
barrier between two neighboring vacua. Due to their relevance to the
electroweak baryogenesis scenario \cite{KRS}, sphalerons were extensively
studied in the literature in extended \textit{SM} variants as in the
\textit{SM} with a singlet \cite{SMS1,SMS2}, the Minimal Supersymmetric
Standard Model \cite{MSSM}; and in the next-to-Minimal Supersymmetric Standard
Model \cite{nMSSM}.

In this work, we will study the sphaleron configuration for a \textit{SU(2)}
gauge-Higgs model in a \textit{5D} background, where the gauge fields
propagate in the \textit{5D} bulk and the Higgs doublet is confined in a
brane. We will focus on the warp effect, by comparing the $AdS_{5}$ results
with the flat geometry case. In the second section, the model is shown, where
the equations of motion (EOM) for the Higgs field and the Kaluza-Klein
(\textit{KK}) gauge modes are given. The sphaleron configuration within this
model is expressed in section three. In the fourth section, we show the
profile functions of the gauge and Higgs fields, as well the values of the
sphaleron energy in different cases. These results will be compared by those
of the \textit{SM}. Finally, we give our conclusion.

\section{SU(2) Gauge Fields in the Bulk}

Let us consider a \textit{SU(2)} Higgs model in the \textit{5D} background
(\ref{5D}), with a general warp factor $a(y)$. The warp factor $a\left(
y\right)  =1$\ refers to the \textit{5D} flat geometry; and $a\left(
y\right)  =e^{-ky}$\ refers to the \textit{AdS}$_{5}$ one. We have $\mu=0,3$
and $M=\mu,5$. In our model, only the gauge fields propagate in the bulk and
the Higgs field is confined in one brane. The action that obeys the symmetry
is%
\begin{equation}
S=\int d^{4}xdy\sqrt{G}\left\{  \mathcal{L}_{bulk}+\Delta(y)\mathcal{L}%
_{brane}\right\}  , \label{S}%
\end{equation}
with $G=\det(g_{MN})$, and $\Delta(y)\equiv2\delta\left(  y\right)
,2\delta\left(  y-\pi R\right)  $\ refers to the Higgs localization in the
\textit{Planck} or \textit{TeV} branes respectively. The boundary Lagrangian
is given by%
\begin{equation}
\mathcal{L}_{brane}=g^{\mu\nu}\left(  D_{\mu}H\right)  ^{\dag}\left(  D_{\nu
}H\right)  -V\left(  H^{\dag}H\right)  ,
\end{equation}
with the covariant derivative%
\begin{equation}
D_{M}H=\left(  \partial_{M}-\frac{i}{2}g_{5}\sigma^{a}A_{M}^{a}\right)  H;
\end{equation}
and $g_{5}=g\sqrt{\pi R}$ is the \textit{5D} \textit{SU(2)} dimensionful gauge
coupling, where $g$ is the \textit{4D} one . The bulk Lagrangian is given by%
\begin{equation}
\mathcal{L}_{bulk}=-\frac{1}{4}g^{MN}g^{QW}F_{MQ}^{a}F_{NW}^{a},
\end{equation}
where the \textit{5D} field strength is given by%
\begin{equation}
F_{MN}^{a}=\partial_{M}A_{N}^{a}-\partial_{N}A_{M}^{a}+g_{5}\epsilon
^{abc}A_{M}^{b}A_{N}^{c}.
\end{equation}
In what follows, we work in the gauge ($\partial^{\mu}A_{\mu}^{a}=0,$
$A_{5}^{a}=0$). The scalar potential has the usual Mexican hat form%
\begin{equation}
V\left(  HH^{\dag}\right)  =\lambda\left(  H^{\dag}H-\upsilon^{2}/2\right)
^{2}, \label{5V}%
\end{equation}
where $\upsilon$\ is the Higgs vev. The equations of motion (EOM) can be
obtained by the vanishing of the action variation, $\delta S=0$, and we get%

\begin{gather}
\Delta(y)a^{4}\left(  y\right)  \left[  g^{\mu\nu}D_{\mu}D_{\nu}%
H+\frac{\partial}{\partial H^{\dag}}V\left(  H^{\dag}H\right)  \right]
=0,\label{heom}\\
\frac{i}{2}g_{5}\Delta(y)a^{4}\left(  y\right)  \left[  H^{\dag}\sigma
^{a}D_{\mu}H-\left(  D_{\mu}H\right)  ^{\dag}\sigma^{a}H\right]  -\partial
_{5}a^{2}\left(  y\right)  \partial_{5}A_{\mu}^{a}+\eta^{\alpha\beta}%
\partial_{\beta}F_{\alpha\mu}^{a}=0, \label{geom}%
\end{gather}

with the boundary condition $\partial_{5}A_{\mu}^{a}=0$\ at both boundaries,
$y=0,\pi R$. The gauge fields have to be factorized using the \textit{KK}
decomposition as%
\begin{equation}
A_{\mu}^{a}\left(  x,y\right)  =\sum\limits_{n}A_{\mu}^{a(n)}\left(  x\right)
\chi^{(n)}(y), \label{kk}%
\end{equation}
with%
\begin{equation}
\int_{0}^{\pi R}\chi^{(n)}(y)\chi^{(m)}(y)dy=\delta_{nm}. \label{gor}%
\end{equation}
Then, the functions $\chi^{(n)}$\ should be the eigenstates of the operator%
\begin{equation}
-\partial_{5}a^{2}\left(  y\right)  \partial_{5}\chi^{(n)}=M_{n}^{2}\chi
^{(n)}, \label{ge}%
\end{equation}
with the condition $\partial_{5}\chi^{(n)}=0$ at both boundaries; $M_{n}$\ are
the \textit{KK} masses. The zero mode $\chi^{(0)}\left(  y\right)
=1/\sqrt{\pi R}$; does not depend on the space-time geometry. In flat
space-time, the heavy modes (\ref{ge}) are given by%
\begin{equation}
\chi^{(n)}\left(  y\right)  =\sqrt{\frac{2}{\pi R}}\cos\left(  \frac{2ny}%
{R}\right)  ,
\end{equation}
with the eigenvalues $M_{n}^{2}=4n^{2}/R^{2}$. However, in the $AdS_{5}$
space-time, they have the form\footnote{This result is given in many works,
like for e.g. \cite{GB1,GB2} and \cite{HQ}.}%
\begin{align}
\chi^{(n)}(y)  &  =\frac{e^{ky}}{a_{n}}\left[  J_{1}\left(  \alpha_{n}%
e^{ky}\right)  -b_{n}Y_{1}\left(  \alpha_{n}e^{ky}\right)  \right]  ,\\
b_{n}  &  =J_{0}\left(  \alpha_{n}\right)  /Y_{0}\left(  \alpha_{n}\right)  ,
\end{align}
with $\alpha_{n}=M_{n}/k$, and $J_{i}$\ and $Y_{i}$\ are the $i-th$ order
Bessel functions of first and second kind, respectively; and $a_{n}$ is a
normalization factor which is computed using (\ref{gor}):
\begin{equation}
a_{n}^{2}=\left.  \frac{e^{2ky}}{2k}\left\{  J_{1}\left(  \alpha_{n}%
e^{ky}\right)  -b_{n}Y_{1}\left(  \alpha_{n}e^{ky}\right)  \right\}
^{2}\right\vert _{y=0}^{y=\pi R}.
\end{equation}
The eigenvalues $M_{n}$ are determined by imposing the boundary condition
$\left.  \partial_{5}\chi^{(n)}=0\right\vert _{y=\pi R}$, which are the zeros
of the quantity
\begin{equation}
Y_{0}\left(  \alpha_{n}e^{\pi kR}\right)  J_{0}\left(  \alpha_{n}\right)
-J_{0}\left(  \alpha_{n}e^{\pi kR}\right)  Y_{0}\left(  \alpha_{n}\right)  .
\label{mevg}%
\end{equation}
These eigenvalues can be obtained numerically.

When inserting (\ref{kk}) in (\ref{S}) and integrating over $y$, we get a
\textit{4D} Lagrangian $\mathcal{L}_{\mathit{4D}}$ as a function of the Higgs
doublet and an infinite number of gauge \textit{KK} modes. The Higgs doublet
is coupled to the \textit{KK} modes through the parameters $\tau_{i} $. In
addition to the quartic couplings between the \textit{KK} modes, which are
characterized by the parameters $\xi_{ijkl}$, there exist also new cubic
couplings characterized by $\gamma_{ijk}$. This feature does exist only in
non-Abelian theories unlike in the Abelian case \cite{GB1,GB2}. The
\textit{4D} Lagrangian is given explicitly in the appendix.

There are some geometry-independent properties of these parameters, like the
invariance under the permutation between each two indices. Also we have the
equalities: $\gamma_{ij0}=\xi_{00ij}=\delta_{ij}$. The \textit{4D SM} can be
recovered by keeping only zero modes in (\ref{L4D}), since all the indices of
zeroth order in (\ref{gpr}) are exactly $1$, whatever the nature of space-time.

The physics at the electroweak scale is more sensitive to the first (and maybe
the second) \textit{KK} mode interactions; therefore, we will give in the
appendix only the numerical values of the coupling of heavy modes with the
first and second \textit{KK} modes. The existence of the warp factor makes a
difference in the masses of the \textit{KK} modes ($M_{i}$) and their
couplings ($\gamma_{ijk}$ and $\xi_{ijkl}$). In what follows, we will
investigate the behavior of the sphaleron configuration with respect to these differences.

\section{Sphaleron Solutions}

It was shown that the \textit{5D} anomaly is independent of the bulk physics;
the cancelation of the \textit{4D} anomaly is sufficient to eliminate the
\textit{5D} one in orbifold theories \cite{Anomaly}. Then the problem of
fermionic current non-conservation can be treated as in a \textit{4D} theory.
In the case of a \textit{5D} fermion coupled to an external gauge potential
$A_{M}^{a}(x,y)$ on an $S^{1}/Z_{2}$ orbifold, the divergent current is given
by \cite{Anomaly}%
\begin{equation}
\partial_{M}\mathbf{J}^{M}(x,y)=\frac{1}{2}[\delta(y)+\delta(y-\pi
R)]F^{a\mu\nu}\tilde{F}_{\mu\nu}^{a}/16, \label{5Current}%
\end{equation}
where $\mathbf{J}^{M}$ is the \textit{5D} fermionic current and $\tilde
{F}_{\mu\nu}^{a}=\frac{1}{2}\epsilon_{\mu\nu\alpha\beta}F^{a\alpha\beta}$ is
the dual field strength. The last term in (\ref{5Current}) represents the
usual \textit{4D} chiral anomaly for a Dirac fermion in an external gauge
potential $A_{M}^{a}(x,y)$. Since the fermions in our model are confined in
one brane, the expression (\ref{5Current}) becomes, after the integration over
the fifth coordinate $y$, like the usual \textit{4D} formula,
\begin{equation}
\partial_{\mu}J^{\mu}(x)=F^{a\mu\nu(0)}\tilde{F}_{\mu\nu}^{a(0)}/32,
\end{equation}
where the label $(0)$ means that only zero modes are taken into account
\cite{Anomaly}; and $J^{\mu}$\ is the \textit{4D} fermionic current. This
means that there is no new contribution to the fermionic currents divergences
beside the \textit{4D} ones. In our model, the Higgs doublet potential on the
brane admits of a minimum, therefore the static energy is bounded from below.
In this case, \textit{N}$_{\mathit{CS}}$\textit{=1/2} represents the so-called
sphaleron configuration \cite{Man0,Man}.

Our system has a \textit{5D} \textit{SU(2)} gauge symmetry; it is invariant
under the gauge transformation%
\begin{equation}
H\rightarrow UH,~i\frac{g}{2}\sigma^{a}A_{M}^{a}\rightarrow i\frac{g}{2}%
\sigma^{a}A_{M}^{a}+\partial_{M}UU^{\dag},
\end{equation}
where $U$ is a \textit{SU(2)} element. In the gauge $A_{5}^{a}=0$, the matrix
\textit{U} should be independent of the fifth dimension; and only the zero
mode will ensure the \textit{SU(2)} gauge invariance. This means that the
sphaleron configuration can be defined for the system ($H$,$~A^{a(0)}$) using
the \textit{4D} transformation matrix $U(\mu,x)$ \cite{Man},
\begin{equation}
U\left(  \mu,x\right)  =\left(
\begin{array}
[c]{cc}%
e^{i\mu}\left(  \cos\mu-i\sin\mu\cos\theta\right)  & e^{i\varphi}\sin\mu
\sin\theta\\
-e^{-i\varphi}\sin\mu\sin\theta & e^{-i\mu}\left(  \cos\mu+i\sin\mu\cos
\theta\right)
\end{array}
\right)  ; \label{U}%
\end{equation}
but this system ($H$,$~A^{a(0)}$) is coupled to the heavy \textit{KK} modes
$A^{a(n\neq0)}$; this effect will be investigated in this work. The sphaleron
configuration can be obtained by making $\mu=\pi/2$.

For reasons of simplicity, we will not use the sphaleron configuration
\cite{Man}, but another, equivalent, representation \cite{Akiba}:
\begin{equation}
H\left(  x\right)  =\frac{\upsilon}{\sqrt{2}}L\left(  r\right)  \left(
\begin{array}
[c]{c}%
0\\
1
\end{array}
\right)  ,~A_{0}^{a}=0,~A_{k}^{a}\left(  x,y\right)  =2\frac{\epsilon
_{akj}x_{j}}{gr^{2}}\sum_{i}\left[  1-f^{(i)}\left(  r\right)  \right]
\chi^{(i)}\left(  y\right)  . \label{Sph5}%
\end{equation}
Here the heavy modes are represented by a similar form as the zero one in
order to make the generalization of the orthogonal gauge $x_{i}A_{i}^{a}=0$
consistent for all the \textit{KK} modes.

Then, when inserting (\ref{Sph5}) in (\ref{heom}) and (\ref{geom}), we get the
differential equations governing the $f^{(i)}(r)$ modes and $L(r)$. The
field's profile functions $L$ and $f^{(i)}$ are given by the solutions of the
system%
\begin{align}
&  \frac{\partial}{\partial\zeta}\zeta^{2}\frac{\partial}{\partial\zeta
}L=2L\sum\limits_{n}\sum\limits_{m}\tau_{n}\tau_{m}\left(  1-f^{(n)}\right)
\left(  1-f^{(m)}\right)  +\frac{\lambda}{2g^{2}}\zeta^{2}L\left(
L^{2}-1\right)  ,\label{sh}\\
&
\begin{array}
[c]{c}%
\zeta^{2}\frac{\partial^{2}}{\partial\zeta^{2}}f^{(i)}=-\frac{\zeta^{2}}%
{4}L^{2}\tau_{i}\sum\limits_{m}\tau_{m}\left(  1-f^{(m)}\right)  -2\left(
1-f^{(i)}\right)  -\zeta^{2}\frac{M_{i}^{2}}{g^{2}\upsilon^{2}}\left(
1-f^{(i)}\right)  +6\sum\limits_{m}\sum\limits_{k}\gamma_{imk}\left(
1-f^{(m)}\right)  \left(  1-f^{(k)}\right)  \\
-4\sum\limits_{m}\sum\limits_{k}\sum\limits_{l}\xi_{imkl}\left(
1-f^{(m)}\right)  \left(  1-f^{(k)}\right)  \left(  1-f^{(l)}\right)
,\label{sg}%
\end{array}
\end{align}
where $\zeta=g\upsilon r$ is the dimensionless radial coordinate, $M_{i}$ are
the \textit{KK} modes eigenmasses; and the $\tau_{i}$ parameters,
$\gamma_{ijk}$\ and $\xi_{ijkl}$\ are given in the appendix. Here, one needs
to mention that the equations (\ref{sh}), (\ref{sg}) and (\ref{en}) are
referring to both cases where the Higgs doublet is localized on the
\textit{Planck} or \textit{TeV} branes. Here one needs to mention that in the
\textit{TeV} brane case, the Higgs doublet as well the \textit{4D} brane
parameters needs to be redefined (for e.g. $a(\pi R)H\rightarrow H$) in order
to be canonically normalized.

The static energy of the system is given by%
\begin{equation}%
\begin{array}
[c]{c}%
E=\frac{4\pi\upsilon}{g}\int_{0}^{\infty}d\zeta\left[  \frac{\zeta^{2}}%
{2}\left(  \frac{\partial}{\partial\zeta}L\right)  ^{2}+\frac{\lambda}{g^{2}%
}\frac{\zeta^{2}}{4}\left(  L^{2}-1\right)  ^{2}+L^{2}\sum\limits_{n}%
\sum\limits_{m}\tau_{n}\tau_{m}\left(  1-f^{(n)}\right)  \left(
1-f^{(m)}\right)  \right. \\
+4\sum\limits_{n}\left\{  \left(  \frac{\partial}{\partial\zeta}%
f^{(n)}\right)  ^{2}+\left[  \frac{2}{\zeta^{2}}+\frac{M_{n}^{2}}%
{g^{2}\upsilon^{2}}\right]  \left(  1-f^{(n)}\right)  ^{2}\right\}  -\frac
{16}{\zeta^{2}}\sum\limits_{n}\sum\limits_{m}\sum\limits_{k}\gamma_{nmk}%
\gamma_{nmk}\left(  1-f^{(m)}\right)  \left(  1-f^{(k)}\right)  \left(
1-f^{(n)}\right) \\
+\left.  \frac{8}{\zeta^{2}}\sum\limits_{n}\sum\limits_{m}\sum\limits_{k}%
\sum\limits_{l}\xi_{nmkl}\left(  1-f^{(n)}\right)  \left(  1-f^{(m)}\right)
\left(  1-f^{(k)}\right)  \left(  1-f^{(l)}\right)  \right]  .
\end{array}
\label{en}%
\end{equation}

When comparing equations (\ref{sh}), (\ref{sg}) and (\ref{en}) with their
corresponding equations in \cite{Man}; we find that instead of the gauge
profile function $f$, we have a summation over an infinite number of $f^{(i)}
$; and also the Higgs-gauge, cubic and quartic gauge-gauge couplings get
modified as%

\begin{equation}%
\begin{array}
[c]{c}%
L^{2}\left(  1-f\right)  ^{2}\rightarrow\sum\limits_{m}\tau_{n}\tau_{m}%
L^{2}\left(  1-f^{(n)}\right)  \left(  1-f^{(m)}\right)  ,\\
\left(  1-f\right)  ^{3}\rightarrow\sum\limits_{m}\sum\limits_{k}\gamma
_{nmk}\left(  1-f^{(n)}\right)  \left(  1-f^{(m)}\right)  \left(
1-f^{(k)}\right)  ,\\
\left(  1-f\right)  ^{4}\rightarrow\sum\limits_{m}\sum\limits_{k}%
\sum\limits_{l}\xi_{nmkl}\left(  1-f^{(n)}\right)  \left(  1-f^{(m)}\right)
\left(  1-f^{(k)}\right)  \left(  1-f^{(l)}\right)  ,
\end{array}
\end{equation}
in addition to the presence of mass terms for non-zero gauge \textit{KK}
modes. Indeed, when neglecting the massive gauge \textit{KK} modes, the EOM
(\ref{sh}) and (\ref{sg}) tend to (11); and (\ref{en}) tends to (10) in
\cite{Man}.

The convergence of the energy functional (\ref{en}) implies the following
boundary conditions on the profiles functions $L$ and $f^{(i)}$.
\begin{align}
\mathit{For}\mathit{~}\zeta &  \rightarrow0:~L\sim\zeta;~~f^{(0)}\sim\zeta
^{2};~~f^{(i)}\sim1,\\
\mathit{and}\mathit{~}\zeta &  \rightarrow\infty:~L\sim1;~f^{(0)}%
\sim1;~f^{(i)}\sim1. \label{fLbc}%
\end{align}
We use the relaxation method to integrate this system of differential
equations. The infinite summations in (\ref{sh}), (\ref{sg}) and (\ref{en})
over the gauge \textit{KK} modes are practically impossible analytically as
well as numerically. We expect that the contributions of the heavy gauge
\textit{KK} modes ($n\geq1$) are just corrections to the energy of the system
($H,$ $A^{a(0)}$); we will consider only a finite number $N$ of the
\textit{KK} modes and then examine the variation the energy (\ref{en}), as
well as the profile functions $L$ and $f^{(n)}$ with respect to this number
$N$ for both cases of flat and warped geometries, with different values of the
warp factor and the first \textit{KK} mass.

\section{Numerical Results and Discussion}

In our computations, we will take the Higgs mass to be around $120$
\textit{GeV}, i.e., $\lambda\simeq0.12$. For a rigorous comparison between the
flat and warped cases, we fix the mass of the first heavy \textit{KK} mode,
which represents in a way the scale of the new physics beyond \textit{SM}, and
we will consider the values $600$ \textit{GeV}, $2$ \textit{TeV} and $10$
\textit{TeV}. In general, the warp factor $w=e^{\pi kR_{w}}$ value is chosen
in a way as to represent the hierarchy between the \textit{Planck} and
\textit{TeV} scales, i.e. $w\sim10^{16}$. But since we are interested also to
investigate its effect on the sphaleron configuration, we will vary the size
of the extra dimension to give it different values for the warp factor:
$w=10^{4}$, $10^{8}$ and the desired one, $10^{16}$.

\begin{figure}[h]
\begin{center}
\resizebox{0.5\textwidth}{7cm}{\includegraphics{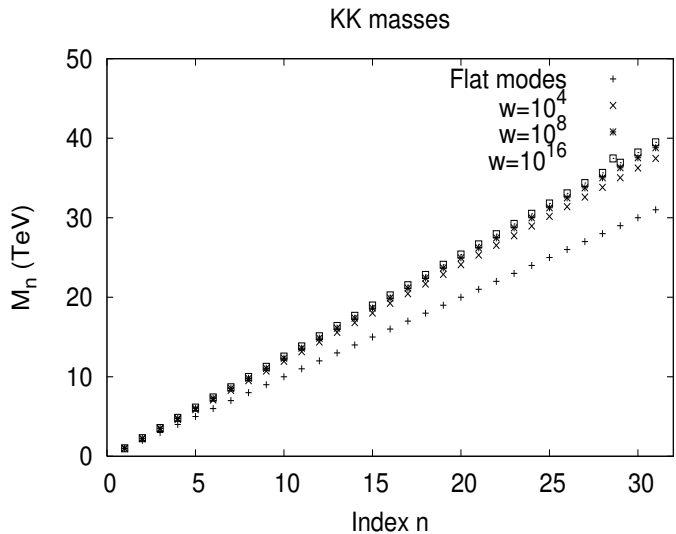}}
\end{center}
\caption{\textit{The masses of the gauge KK modes for the cases of flat and
warped geometry with different values of the warp factor w.}}%
\label{KKMs}%
\end{figure}

In Fig. \ref{KKMs}, the masses of \textit{KK} modes are shown for both flat
and warped backgrounds, where the first \textit{KK} heavy mode mass is chosen
to be $1$ \textit{TeV}. It is clear that the flat modes are just multipliers
of the first heavy one, while the existence of the warp factor makes the
warped mode masses increasing with respect to the warp factor $w$.

For the Higgs-gauge and gauge-gauge couplings, they are given in unit of the
\textit{SU(2)} coupling $g$; by the parameters $\tau$, $\gamma$\ and $\xi$.
All these parameters are of order $\mathcal{O}(1)$ in the flat geometry. In
warped geometry, the situation is different, the $\tau$ parameters; that
represent the couplings of the Higgs with gauge \textit{KK} modes, depend on
which boundary the Higgs filed is located in. If the Higgs field is located in
the \textit{Planck} brane, these parameters are negative and their modulus is
less than unity and decaying with respect to the \textit{KK} masses, and also
with respect to the warp effect. If the Higgs doublet is located in the
\textit{TeV} brane, the values of the $\tau$\ parameters are of the order
$\mathcal{O}(1)$ but positive for odd modes and negative for the even ones;
and their modules are almost stable with respect to the \textit{KK} masses.
The previous difference between the two cases will not change significantly
the profile functions of $L$ and $f^{(i)}$ or the sphaleron energy (\ref{en}).
The difference between the sphaleron energy in both cases is less than
$0.004$\% for $w=10^{16}$\ and $M_{1}=1$ \textit{TeV}.

The $\gamma$ parameters that describe the cubic couplings between the gauge
\textit{KK} modes are also small in the $AdS_{5}$ background and decaying with
respect to the \textit{KK} masses. However, the $\xi$ parameters that
represent the quartic couplings between the gauge \textit{KK} modes are large
(for e.g. $\xi_{1,1,1,1}\sim46$) and decaying with respect to the \textit{KK}
masses but still remaining large (for e.g. $\xi_{30,30,30,30}\sim27$).

\begin{figure}[h]
\begin{center}
\resizebox{0.5\textwidth}{5.5cm}{\includegraphics{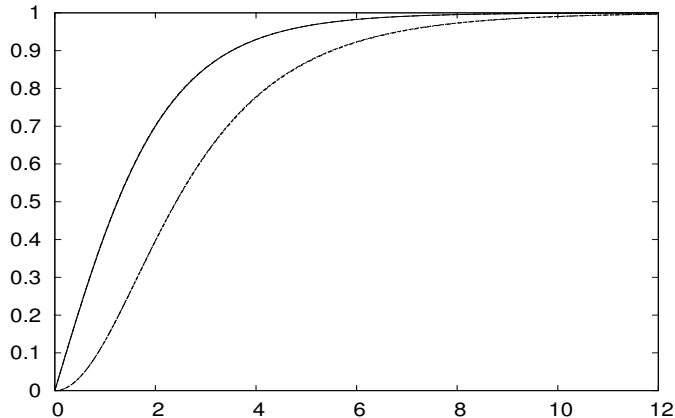}}
\end{center}
\caption{\textit{The profile functions }$L$\textit{ (upper curve) and
}$f^{(0)}$\textit{ (lower curve) for the SM case, flat geometry and the warped
geometry as a function of the dimensionless radial coordinate }$\zeta
$\textit{. Each profile function is almost identical for the different cases.
This plot was performed taking into account the first 10 heavy KK modes for
both flat and warped geometries for }$M_{1}=1$\textit{ TeV and }$w=10^{16}%
$\textit{.}}%
\label{pro}%
\end{figure}

\begin{figure}[h]
\begin{center}
\resizebox{0.45\textwidth}{5cm}{\includegraphics{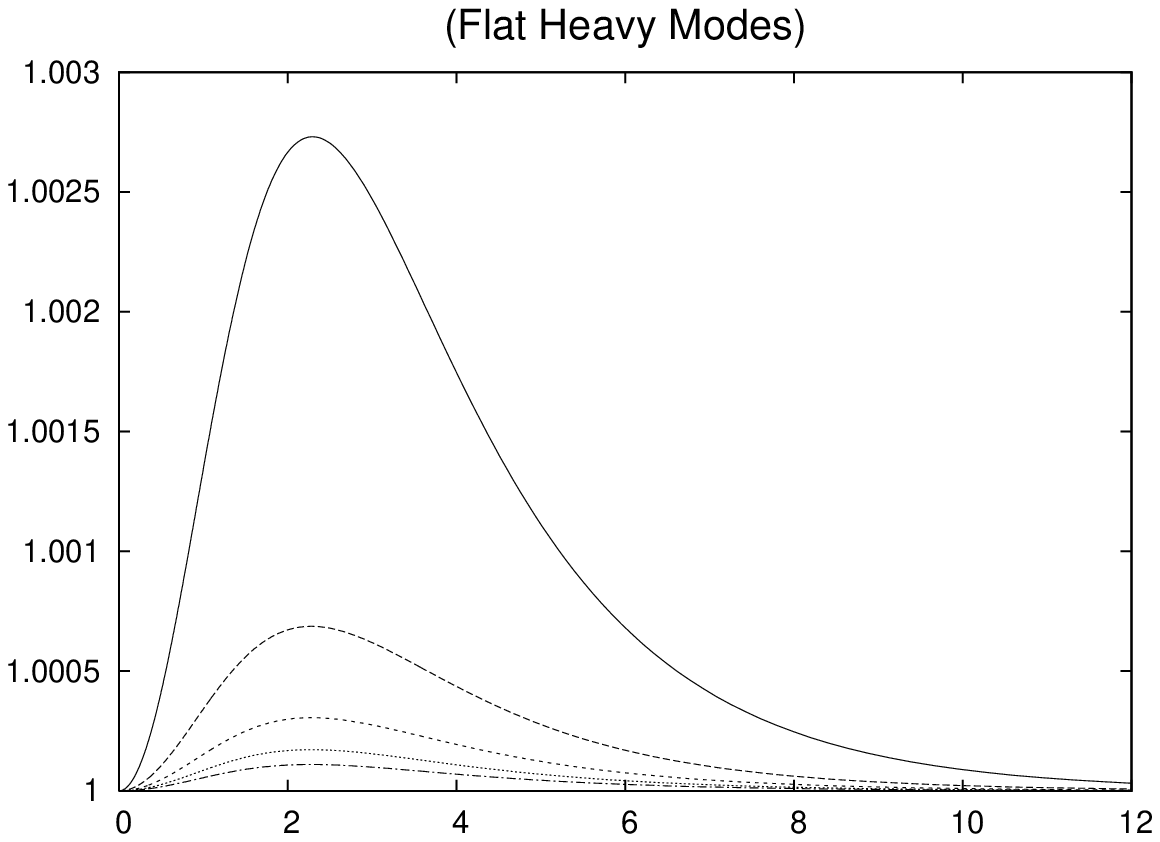}}
\resizebox{0.45\textwidth}{5cm}{\includegraphics{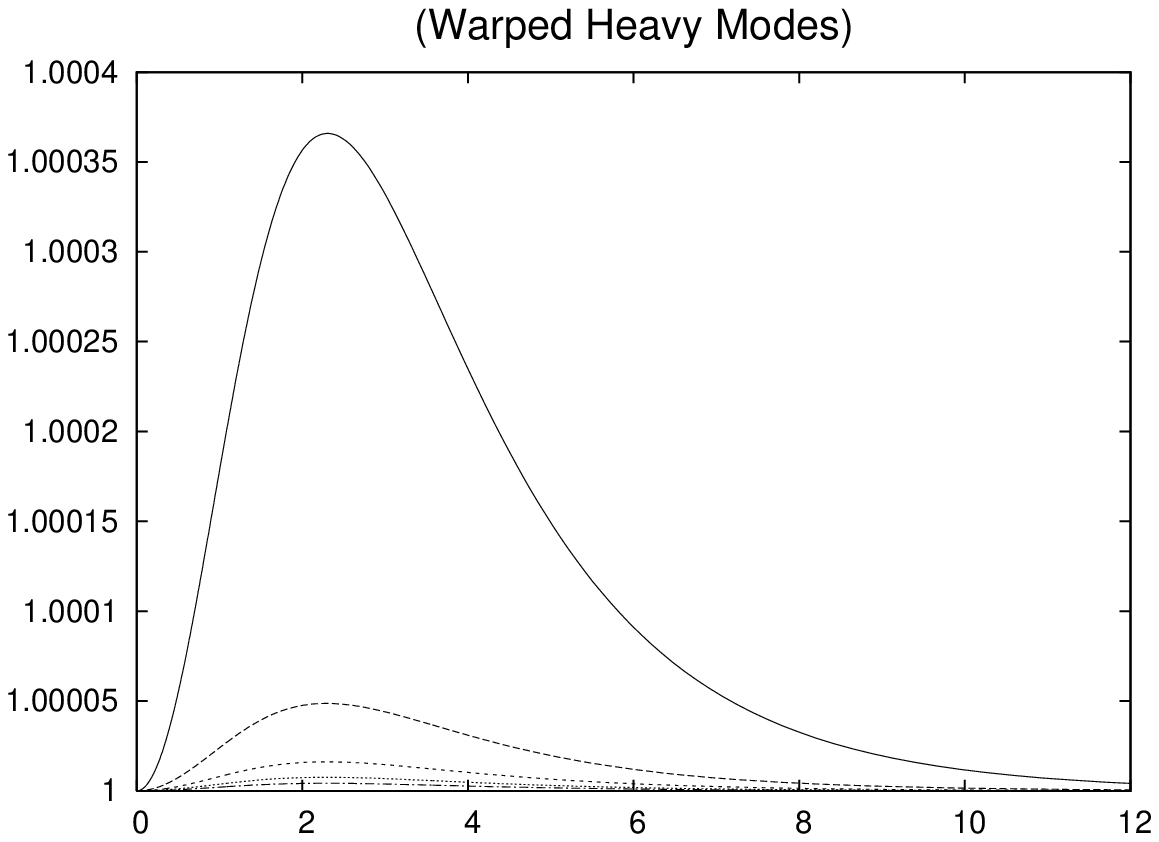}}
\end{center}
\caption{\textit{From up to down, here are the profile functions }$f^{(i)}%
$\textit{, of the first five heavy modes for the flat geometry case (up) and
warped geometry (down) for the same values of }$M_{1}$\textit{ and }%
$w$\textit{ taken in Fig. \ref{pro}, as a function of the dimensionless radial
coordinate }$\zeta$\textit{.}}%
\label{hpro}%
\end{figure}

The profile functions $L$ and $f^{(0)}$ are given in Fig. \ref{pro}. They are
very close to the \textit{SM} ones to a very high precision for both the cases
of flat and warped geometries. This feature does not depend on $N$, the number
of the heavy modes taken into account to solve (\ref{sh}) and (\ref{sg}).
However, the profile functions of the heavy modes $f^{(i)}$, as shown in Fig.
\ref{hpro}, are just deviations from 1; and these deviations decrease with
respect to the \textit{KK} masses.

We remark that the profile functions of the heavy modes $f^{(i)}$, are more
suppressed in the case of warped geometry than in the flat one. However, the
suppression effect decreases if we decrease the warp factor; for, e.g., when
taking the warp factor to be $w=10^{4}$ instead of $10^{16}$, the maximum of
$f^{(1)}$ (the upper curve in the right side of Fig. \ref{hpro}) increases
from $1.00037$ to $1.00075$. This suppression increases also if we increase
the first \textit{KK} mode mass.

Due to the fact that the profile functions of $L$ and $f^{(0)}$ practically do
not change with respect the \textit{SM} results, and in addition to the
suppression of the heavy modes profile functions, one expects that the
sphaleron energy should not be very different from the \textit{SM} value, but
this is not guaranteed due to the infinite number of terms in Eq. (\ref{en}),
as well the increasing \textit{KK} mass values, unless confirmed numerically.

\begin{figure}[h]
\begin{center}
\resizebox{0.48\textwidth}{6cm}{\includegraphics{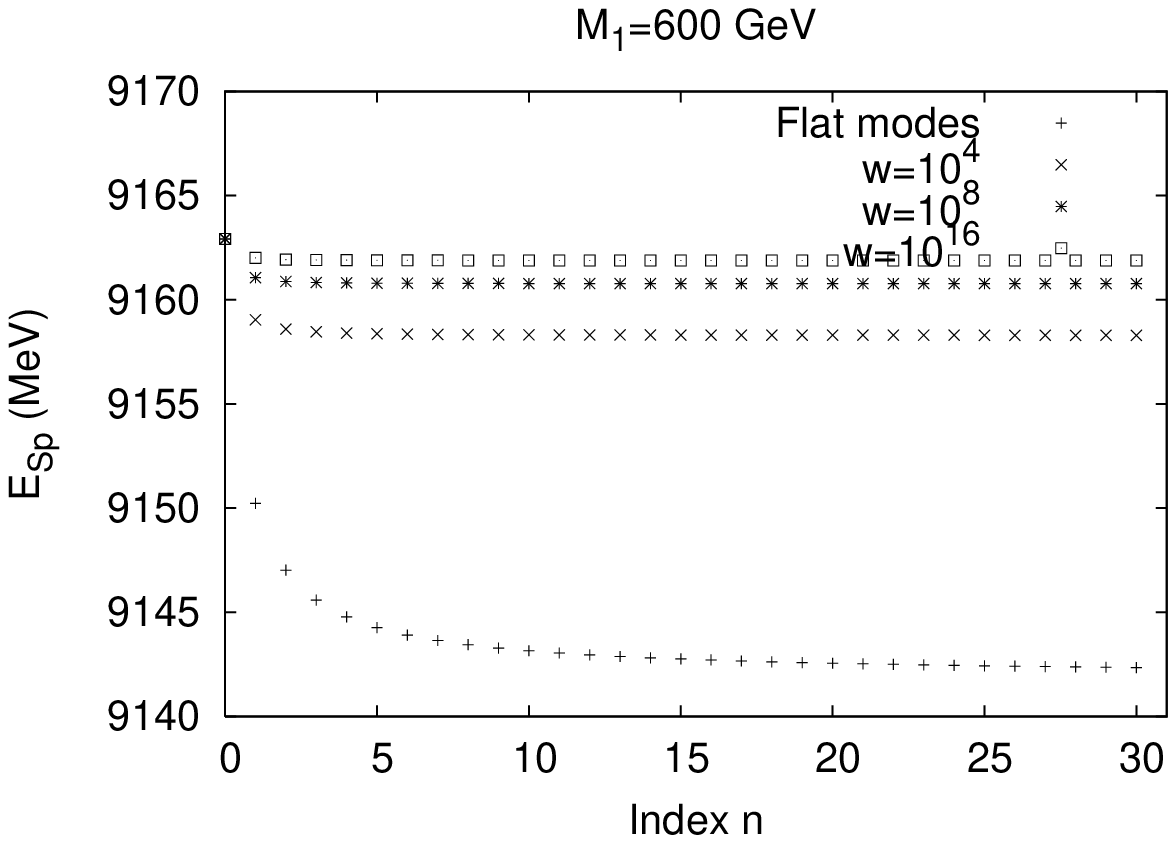}}
\resizebox{0.48\textwidth}{6cm}{\includegraphics{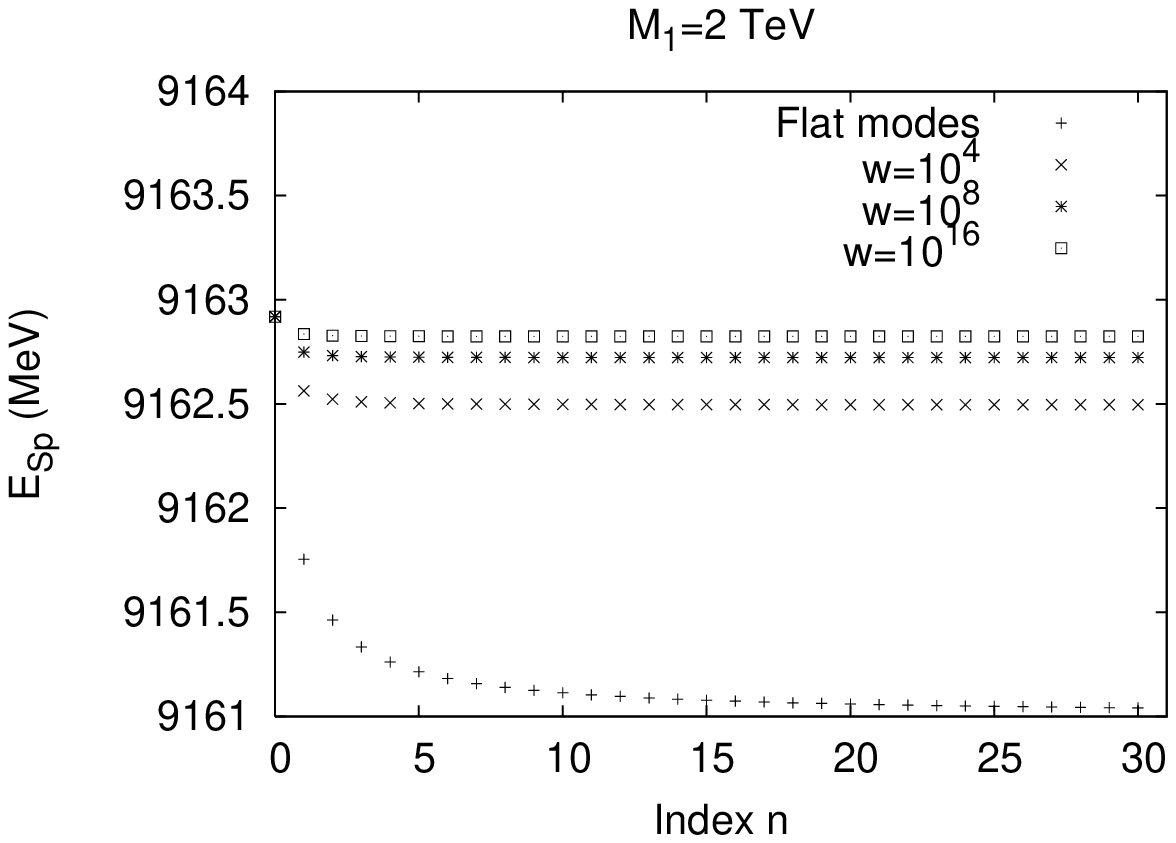}}
\resizebox{0.48\textwidth}{6cm}{\includegraphics{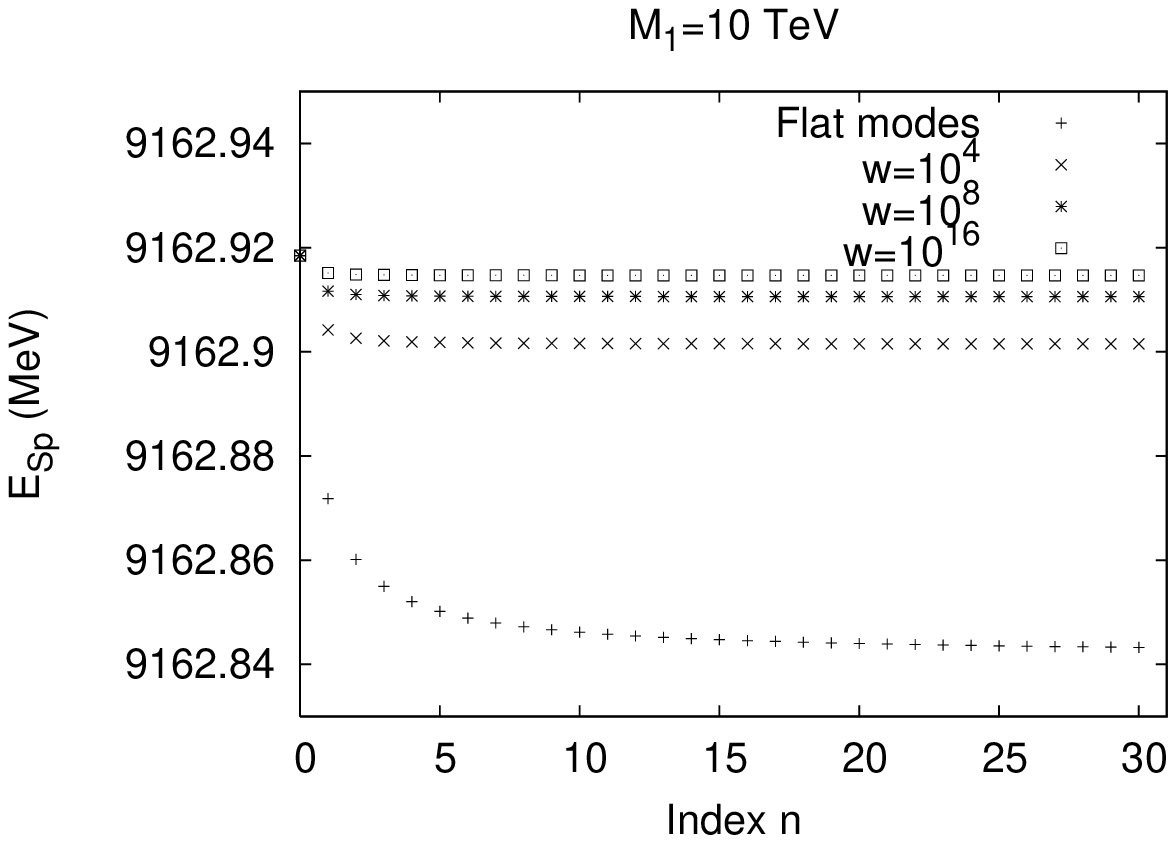}}
\end{center}
\caption{\textit{The dependence of the sphaleron energy on the number of heavy
KK modes that are taken into account to estimate (\ref{en}); for different
values of the first KK mode mass.}}%
\label{sph}%
\end{figure}

To check this, we compute the sphaleron energy (\ref{en}) taking into account
a finite number $N$ of \textit{KK} modes for the different values of the first
heavy \textit{KK} mode mass and the warped factor mentioned above. The
sphaleron energy dependence on the index $N$ is shown in Fig. \ref{sph}.

The first remark on the results in Fig. \ref{sph}; is that the sphaleron
energy does differ significantly from the \textit{SM} value; its largest
deviation is in the case of a small mass of the first \textit{KK} heavy mode
with flat geometry (first plot in Fig. \ref{sph}), which is $-0.06~\%$, i.e.
much less than $1~\%$. Also, the existence and largeness of the warp factor
makes the sphaleron energy practically identical to the \textit{SM} value.
However, this feature is due to the sphaleron configuration itself, i.e.
$(H,A_{\mu}^{(0)})$, rather than the decoupling effect of the heavy
\textit{KK} modes, because if we consider an extreme case of a flat geometry
with a small mass for the first \textit{KK} mode (for e.g. $300$ \textit{GeV},
and then $100$ \textit{GeV}), the sphaleron energy decreases only by $-0.9~\%$
and $-6~\%$, respectively. This can be explained by the fact that most of the
sphaleron energy is coming from the contributions of the gauge zero mode and
the Higgs fields; and the profile functions of these fields are determined by
self-interactions as well as interactions with each other rather than their
interactions with the heavy \textit{KK} modes. Then one can say that the heavy
\textit{KK} modes are just compensating fields in the EOM (\ref{sh}) and
(\ref{sg}), as in the case of the singlet in the model of \textit{SM}+singlet
\cite{SMS2}. This could explain the fact that the contributions of the
\textit{KK} modes to the sphaleron energy (\ref{en}) are very small even
though their cubic ($\gamma$) and quartic ($\xi$) coupling are (very) large.
Indeed, the sphaleron energy (\ref{en}) is more sensitive to the first
\textit{KK} eigenmass rather than to the couplings $\tau$, $\gamma$ and $\xi$.

At finite temperature, we do not expect to have a deviation in sphaleron
field's profile functions as well as in the values of sphaleron energy from
the results of the \textit{SM} \cite{Tspb}; and the \textit{B+L} anomaly is
almost the same as in the standard theory. Then the criterion for a strongly
first-order phase transition remains the known one, $\upsilon_{c}/T_{c}\geq1 $
\cite{PT}.

\section{Conclusion}

In this work, the sphaleron configuration for a Higgs model in a \textit{5D}
space-time is studied, where the Higgs is confined in a brane and the gauge
field resides in the \textit{5D} bulk. When we made the \textit{KK}
decomposition of the gauge field, we found that possible interactions (cubic
and quartic) between different \textit{KK} modes are possible due to the
non-Abelian nature of the symmetry group unlike the Abelian case
\cite{GB1,GB2}. The strength of these interactions depends on the space-time
nature. The strength of the interaction with the Higgs doublet depends on
where it is located in.

We defined the sphaleron configuration in this case, where we got the
equations like the \textit{SM} case, but corrected by the existence of the
\textit{KK} heavy modes. Practically the profile functions of the Higgs and
zero mode gauge fields do not change when comparing with the \textit{SM}
results; and the heavy mode profile functions are just little deviations from
1. The suppression of this deviation from unity is proportional to the
\textit{KK} order. Also the existence of a strong warp factor (like
$w=10^{16}$) suppresses these deviations by one order of magnitude.

We checked also that the sphaleron energy has the same value as the
\textit{SM} one. The heavy \textit{KK} modes do not practically contribute to
the sphaleron energy; and their presence decreases the value of sphaleron
energy by $-0.25\%$ for a light mass of the first \textit{KK} heavy mode (600
\textit{GeV}) in a flat geometry. The existence of a warp factor; or the
increasing of the mass of the first \textit{KK} heavy mode, which represents
somehow the new physics scale, suppresses the deviation from the \textit{SM} results.

This allows us to suppose that at finite temperature, the previous
results should differ from those of the \textit{SM. }In addition to
the fact that the \textit{5D} \textit{B+L} anomaly is identical to
the \textit{4D} one, the criterion of a strong first-order phase
transition, $\upsilon_{c}/T_{c}\geq1 $, is still valid for these
models.

\vspace{1cm} \textbf{Acknowledgements}: \textit{I want to thank
Mikko Laine for his useful comments as well for the warm hospitality
at Bielefeld University. This work was supported by both the German
Academic Exchange Service (DAAD) and the Algerian Ministry of Higher
Education and Scientific Research under the cnepru-project
D0092007148.}

\appendix

\section{Explicit 4D Lagrangian}

The \textit{4D} theory can be obtained by integrating over the fifth
dimension. Here we explicitly give the \textit{4D} Lagrangian with its
different parameters that describe the couplings of the gauge \textit{KK}
modes with themselves as well as with the Higgs doublet. It is given by
\begin{equation}%
\begin{array}
[c]{l}%
\mathcal{L}_{\mathit{4D}}=\eta^{\mu\nu}\partial_{\mu}H^{\dag}\partial_{\nu
}H-V\left(  H^{\dag}H\right)  -\frac{i}{2}g\eta^{\mu\nu}\left[  \partial_{\nu
}H^{\dag}\sigma^{a}H-H^{\dag}\sigma^{a}\partial_{\nu}H\right]  \sum
\limits_{n}\tau_{n}A_{\mu}^{a(n)}+\frac{1}{2}\eta^{\mu\nu}\sum\limits_{n}%
\sum\limits_{m}(\tau_{n}\tau_{m}\frac{g^{2}}{2}H^{\dag}H\\
+\delta_{nm}M_{n}^{2})A_{\mu}^{a(n)}A_{\nu}^{a(m)}-\frac{1}{2}\eta^{\mu\nu
}\eta^{\alpha\beta}\sum\limits_{n}\left[  \partial_{\mu}A_{\alpha}%
^{a(n)}\partial_{\nu}A_{\beta}^{a(n)}-\partial_{\alpha}A_{\mu}^{a(n)}%
\partial_{\nu}A_{\beta}^{a(n)}\right]  -g\eta^{\mu\nu}\eta^{\alpha\beta
}\epsilon^{abc}\sum\limits_{n}\sum\limits_{m}\sum\limits_{k}\gamma_{nmk}%
\times\\
A_{\nu}^{b(m)}A_{\beta}^{c(k)}\partial_{\mu}A_{\alpha}^{a(n)}-\frac{g^{2}}%
{4}\eta^{\mu\nu}\eta^{\alpha\beta}\epsilon^{abc}\epsilon^{ade}\sum
\limits_{n}\sum\limits_{m}\sum\limits_{k}\sum\limits_{l}\xi_{nmkl}A_{\mu
}^{b(n)}A_{\alpha}^{c(m)}A_{\nu}^{d(k)}A_{\beta}^{e(l)}.
\end{array}
\label{L4D}%
\end{equation}

The parameters $\tau_{n}$, $\gamma_{nmk}$\ and $\xi_{nmkl}$\ are given by%
\begin{equation}%
\begin{array}
[c]{c}%
\tau_{n}=\sqrt{\pi R}\int\limits_{0}^{\pi R}\sqrt{G}\mathbf{\Delta}\left(
y\right)  \chi^{(n)}\left(  y\right)  dy,~\gamma_{nmk}=\sqrt{\pi R}%
\int\limits_{0}^{\pi R}dy\chi^{(n)}\left(  y\right)  \chi^{(m)}\left(
y\right)  \chi^{(k)}\left(  y\right)  ,\\
\xi_{nmkl}=\pi R\int\limits_{0}^{\pi R}dy\chi^{(n)}\left(  y\right)
\chi^{(m)}\left(  y\right)  \chi^{(k)}\left(  y\right)  \chi^{(l)}\left(
y\right)  .
\end{array}
\label{gpr}%
\end{equation}

In a flat space-time, these parameters can be reduced to%
\begin{equation}%
\begin{array}
[c]{l}%
\tau_{n}=1/\sqrt{2},~\gamma_{nmk}=\left\{  \delta_{0,m+k-n}+\delta
_{0,m-k-n}+\delta_{0,m-k+n}\right\}  /\sqrt{2},\\
\xi_{nmkl}=\left\{  \delta_{0,n+m-k-l}+\delta_{0,n+m+k-l}+\delta
_{0,n+m-k+l}+\delta_{0,n-m+k+l}+\delta_{0,n-m-k-l}\right. \\
\left.  +\delta_{0,n-m+k-l}+\delta_{0,n-m-k+l}\right\}  /2.
\end{array}
\end{equation}
In the $AdS_{5}$ space-time, the formulae of the $\tau_{i}$ parameters are
given in both the cases where Higgs field is confined in the \textit{Planck}
(Pl) and \textit{TeV} branes by%
\begin{equation}
\tau_{n}^{(Pl)}=\sqrt{\pi R}\chi^{(n)}(0),~\tau_{n}^{(\mathit{TeV})}=\sqrt{\pi
R}\chi^{(n)}(\pi R). \label{tau}%
\end{equation}
In the following table, we give the first 10 values of the $\tau_{i}%
$\ parameters for different values of the warp factor. \begin{table}[h]
\begin{center}
$%
\begin{tabular}
[c]{c|ccc|}\cline{2-4}
&  & $\tau_{n}^{(Pl)}$ & \\\hline
\multicolumn{1}{|c|}{i} & $w=10^{4}$ & \multicolumn{1}{|c}{$w=10^{8}$} &
\multicolumn{1}{|c|}{$w=10^{16}$}\\\hline
\multicolumn{1}{|c|}{1} & -0.1955 & \multicolumn{1}{|c}{-0.1352} &
\multicolumn{1}{|c|}{-0.0945}\\
\multicolumn{1}{|c|}{2} & -0.1453 & \multicolumn{1}{|c}{-0.0950} &
\multicolumn{1}{|c|}{-0.0645}\\
\multicolumn{1}{|c|}{3} & -0.1236 & \multicolumn{1}{|c}{-0.0782} &
\multicolumn{1}{|c|}{-0.0523}\\
\multicolumn{1}{|c|}{4} & -0.1107 & \multicolumn{1}{|c}{-0.0683} &
\multicolumn{1}{|c|}{-0.0453}\\
\multicolumn{1}{|c|}{5} & -0.1018 & \multicolumn{1}{|c}{-0.0617} &
\multicolumn{1}{|c|}{-0.0405}\\
\multicolumn{1}{|c|}{6} & -0.0952 & \multicolumn{1}{|c}{-0.0568} &
\multicolumn{1}{|c|}{-0.0371}\\
\multicolumn{1}{|c|}{7} & -0.0900 & \multicolumn{1}{|c}{-0.0530} &
\multicolumn{1}{|c|}{-0.0344}\\
\multicolumn{1}{|c|}{8} & -0.0858 & \multicolumn{1}{|c}{-0.0500} &
\multicolumn{1}{|c|}{-0.0322}\\
\multicolumn{1}{|c|}{9} & -0.0823 & \multicolumn{1}{|c}{-0.0473} &
\multicolumn{1}{|c|}{-0.0304}\\
\multicolumn{1}{|c|}{10} & -0.07936 & \multicolumn{1}{|c}{-0.0452} &
\multicolumn{1}{|c|}{-0.0289}\\\hline
\end{tabular}%
\begin{tabular}
[c]{||c|c|c|}\cline{1-3}%
\multicolumn{1}{||c|}{} & $\tau_{n}^{(TeV)}$ & \\\hline
$w=10^{4}$ & $w=10^{8}$ & $w=10^{16}$\\\hline
2.1549 & 3.0379 & 4.2930\\
-2.1509 & -3.0363 & -4.2924\\
2.1495 & 3.0359 & 4.2923\\
-2.1488 & -3.0356 & -4.2922\\
2.1484 & 3.0355 & 4.2921\\
-2.1481 & -3.0354 & -4.2921\\
2.1479 & 3.0353 & 4.2921\\
-2.1477 & -3.0353 & -4.2921\\
2.1475 & 3.0352 & 4.2920\\
-2.1474 & -3.0352 & -4.2920\\\hline
\end{tabular}
$
\end{center}
\caption{\textit{Different values of the parameters $\tau_{i}$ for different
values of the warp factor in both the cases where the Higgs doublet is
confined in the Planck brane (left) or TeV brane (right).}}%
\end{table}

For the parameters $\gamma$\ and $\xi$, it is easy to check that they depend
only on the warp factor $w$, and not on the first \textit{KK} mass $M_{1}$.
Their formulae are complicated; and therefore they could be computed numerically.

As stated above in section 2, it is important to estimate the couplings of the
heavy modes with the zero and first one (and maybe the second one). Here we
give the numerical values of$~\gamma_{1,1,i}$, which represents the cubic
coupling of two one modes with a heavier one ($i\geq2$), or equivalently, the
quartic coupling of a zero mode, two one modes and a heavier one. We give also
the value of $\xi_{1,1,1,i}$, which represents the quartic coupling of three
one modes and a heavier one, taking the value of the warp factor to be
$w=10^{4},~10^{8},~10^{16}$. \begin{table}[h]
\begin{center}
$%
\begin{array}
[c]{c}%
w=10^{4}\\%
\begin{tabular}
[c]{|c|c|c||c|c|}\hline
i & $\gamma_{1,1,i}$ & $\xi_{1,1,1,i}$ & $\gamma_{1,2,i}$ & $\xi_{1,1,2,i}%
$\\\hline
1 & 2.9616 & 10.9652 & -1.0925 & -5.4800\\\hline
2 & -1.0925 & -5.4800 & 2.0279 & 6.6308\\\hline
3 & 0.0253 & 1.3800 & -1.1505 & -4.6203\\\hline
4 & 5.78$\times10^{-4}$ & -0.0676 & 0.0364 & 1.4808\\\hline
5 & 9.46$\times10^{-4}$ & 3.95$\times10^{-3}$ & 7.45$\times10^{-4}$ &
-0.0895\\\hline
6 & 1.49$\times10^{-4}$ & -2.51$\times10^{-3}$ & 1.83$\times10^{-3}$ &
6.39$\times10^{-3}$\\\hline
7 & 1.48$\times10^{-4}$ & 2.96$\times10^{-4}$ & 2.73$\times10^{-4}$ &
-4.23$\times10^{-3}$\\\hline
8 & 4.32$\times10^{-5}$ & -3.59$\times10^{-4}$ & 3.33$\times10^{-4}$ &
6.06$\times10^{-4}$\\\hline
9 & 4.01$\times10^{-5}$ & 4.54$\times10^{-5}$ & 9.21$\times10^{-5}$ &
-6.90$\times10^{-4}$\\\hline
10 & 1.60$\times10^{-5}$ & -8.83$\times10^{-5}$ & 9.86$\times10^{-5}$ &
1.12$\times10^{-4}$\\\hline
\end{tabular}
\\
w=10^{8}\\%
\begin{tabular}
[c]{|c|c|c||c|c|}\hline
i & $\gamma_{1,1,i}$ & $\xi_{1,1,1,i}$ & $\gamma_{1,2,i}$ & $\xi_{1,1,2,i}%
$\\\hline
1 & 4.4339 & 22.7671 & -1.4517 & -10.8886\\\hline
2 & -1.4517 & -10.8886 & 3.0403 & 13.7351\\\hline
3 & 0.0250 & 2.4249 & -1.5438 & -9.2595\\\hline
4 & -4.39$\times10^{-4}$ & -0.0966 & 0.0365 & 2.6336\\\hline
5 & 8.48$\times10^{-4}$ & 9.39$\times10^{-3}$ & -1.11$\times10^{-3}$ &
-0.1304\\\hline
6 & 5.61$\times10^{-5}$ & -3.62$\times10^{-3}$ & 1.68$\times10^{-3}$ &
0.0152\\\hline
7 & 1.27$\times10^{-4}$ & 7.90$\times10^{-4}$ & 6.61$\times10^{-5}$ &
-6.29$\times10^{-3}$\\\hline
8 & 2.41$\times10^{-5}$ & -4.99$\times10^{-4}$ & 2.91$\times10^{-4}$ &
1.57$\times10^{-3}$\\\hline
9 & 3.35$\times10^{-5}$ & 1.43$\times10^{-4}$ & 4.43$\times10^{-5}$ &
-1.00$\times10^{-3}$\\\hline
10 & 1.01$\times10^{-5}$ & -1.18$\times10^{-4}$ & 8.39$\times10^{-5}$ &
3.23$\times10^{-4}$\\\hline
\end{tabular}
\\
w=10^{16}\\%
\begin{tabular}
[c]{|c|c|c||c|c|}\hline
i & $\gamma_{1,1,i}$ & $\xi_{1,1,1,i}$ & $\gamma_{1,2,i}$ & $\xi_{1,1,2,i}%
$\\\hline
1 & 6.4532 & 46.5505 & -1.990 & -21.6920\\\hline
2 & -1.9899 & -21.6920 & 4.4233 & 28.0336\\\hline
3 & 0.0308 & 4.5373 & -2.1258 & -18.5267\\\hline
4 & 8.10$\times10^{-4}$ & -0.1620 & 0.0431 & 4.9618\\\hline
5 & 2.75$\times10^{-3}$ & 0.0193 & -1.68$\times10^{-3}$ & -0.2216\\\hline
6 & 1.68$\times10^{-3}$ & -6.44$\times10^{-3}$ & 3.00$\times10^{-3}$ &
0.0317\\\hline
7 & 1.64$\times10^{-3}$ & 1.41$\times10^{-3}$ & 1.13$\times10^{-3}$ &
-0.0110\\\hline
8 & 1.33$\times10^{-3}$ & -1.09$\times10^{-3}$ & 1.36$\times10^{-3}$ &
3.11$\times10^{-3}$\\\hline
9 & 0.0988 & 2.94$\times10^{-4}$ & -1.82$\times10^{-5}$ & -1.73$\times10^{-3}%
$\\\hline
10 & 1.01$\times10^{-3}$ & -3.98$\times10^{-4}$ & 5.29$\times10^{-5}$ &
7.00$\times10^{-4} $\\\hline
\end{tabular}
\end{array}
$
\end{center}
\caption{\textit{Different values of the cubic ($\gamma_{1,1,i}$ and
$\gamma_{1,2,i}$) and quartic ($\xi_{1,1,1,i}$ and $\xi_{1,1,2,i}$)
gauge-gauge couplings for $w=10^{4}$, $10^{8}$, $10^{16}$.}}%
\end{table}

\end{document}